# The Economics of BitCoin Price Formation[1]

Pavel Ciaian[1,2,3], Miroslava Rajcaniova[2,3,4], d'Artis Kancs[1,2,3]

[1]European Commission (DG Joint Research Centre)
[2]Economics and Econometrics Research Institute (EERI)
[3]Catholic University of Leuven (LICOS)
[4]Slovak University of Agriculture in Nitra (SUA)

**Abstract**

This paper analyses the relationship between BitCoin price and supply-demand fundamentals of BitCoin, global macro-financial indicators and BitCoin's attractiveness for investors. Using daily data for the period 2009-2014 and applying time-series analytical mechanisms, we find that BitCoin market fundamentals and BitCoin's attractiveness for investors have a significant impact on BitCoin price. Our estimates do not support previous findings that the macro-financial developments are driving BitCoin price.

[1] The authors are grateful to Tony Tam from for providing access to the BitCoin data of Bitcoinpulse. The views expressed are purely those of the authors and may not in any circumstances be regarded as stating an official position of the European Commission.

# 1 Introduction

Over the last few years, a wide range of digital currencies, such as BitCoin, LiteCoin, PeerCoin, AuroraCoin, DogeCoin and Ripple, have emerged. The most prominent among them is BitCoin, both in terms of an impressive price development and market capitalisation. Its price increased from zero value at the time of its inception in 2009 to around $13 per a BitCoin in January 2013, and subsequently shot up by more than 8000% to around $1100 at the end of 2013 (see Figure 1). In March 2014 the total market capitalisation of BitCoin was more than $5.6 billion.

The rise of BitCoin's popularity has attracted a growing interest among economists in general (e.g. Grinberg 2011; Barber et al. 2012; Kroll, Davey and Felten 2013; Moore and Christin 2013), and in BitCoin' price formation in particular (e.g. Buchholz et al. 2012; Kristoufek 2013; van Wijk 2013). Several factors affecting BitCoin price have been identified in the previous literature: (i) market fundamentals such as BitCoin supply and demand (Buchholz et al. 2012); (ii) attractiveness for investors (Kristoufek 2013); and (iii) development of global financial indicators (van Wijk 2013).

Buchholz et al. (2012) note that an important determinant of BitCoin price (as price of any currency) is the interaction between BitCoins' supply and demand. The supply of BitCoin determines the amount of units in circulation and thus its scarcity on the market. The demand of BitCoin is mainly determined by transaction demand as a medium of exchange. Buchholz et al. argue that BitCoin price is an outcome of interaction between supply and demand.

According to Kristoufek (2013), the price formation of BitCoin cannot be explained by standard economic theories,[2] because supply-demand fundamentals, which usually form the basis of currency price formation, are absent on BitCoin markets. First, BitCoin is not issued by a specific central bank or government and thus is detached from the real economy. Second, the demand (and supply) for BitCoin is driven also by investors' speculative behaviour, because there is no interest rate for the digital currencies and thus profits can be earned only from price changes.

Van Wijk (2013) stresses the role of global financial development, captured e.g. by stock exchange indices, exchange rates, and oil prices measures in determining BitCoin price. Van Wijk finds evidence that the Dow Jones index, the euro-dollar exchange rate, and oil price have a significant impact on the value of BitCoin in the long run.

An important shortcoming of previous studies is that they study the impact of each BitCoin's price determinant separately, hence they do not consider interactions between them. The present paper attempts to close this research gap by accounting of all three types of BitCoin price determinants identified in the previous literature: supply-demand fundamentals,

---

[2] For example, future cash-flows model, purchasing power parity, or uncovered interest rate parity.

investors' behaviour and global financial indicators to explain the formation of BitCoin price, and to account for their interactions

Understanding the BitCoin's price formation is highly relevant both from a general monetary policy point of view and from a BitCoin's ability to serve as a medium of exchange for global economy's point of view.[3]

## 2 The rise of BitCoin

BitCoin is a peer-to-peer payment system created in 2009. It is the first open source digital currency, and BitCoin is managed by an open source *software algorithm* that uses the global internet network both to create the BitCoins as well as to record and verify transactions. Being a cryptocurrency, BitCoin uses the principles of cryptography to control the creation and transfer of money. Access to the BitCoin network requires downloading the BitCoin software on personal computer and joining the BitCoin network, which allows participants to engage in operations, and update and verify the transactions.

Compared to a standard fiat currency, such as dollars or euros, the key distinguishing feature of BitCoin is that the quantity of units in circulation is not controlled by a person, group, company, central authority, or government, but a software algorithm controls the amount of BitCoins issued.[4] A fixed amount of BitCoins is issued at a fixed a-priori defined and publicly known rate, implying that the stock of BitCoins increases at a decreasing rate. In 2140 the BitCoin growth rate will converge to zero, when the maximum amount of BitCoins in circulation will reach 21 million units. Hence, the maximum stock of BitCoins will not change after 2140.

BitCoins can be used to buy goods or services worldwide, provided that transaction partners accept BitCoin as a mean of payment. A transaction implies that BitCoin owners transfer their ownership of a certain number of BitCoins, in exchange for goods and services. An increasing number of companies accept BitCoins as payments for their goods and services (CoinDesk 2014). BitCoins can be also exchanged for other currencies.

To summarise, BitCoin is a fiat currency without an intrinsic value. In contrast to standard government backed fiat currencies, e.g. dollar, euro, BitCoin is developed outside of an underlying economy or issuing institution, implying that there are no macroeconomic fundamentals that would determine its price formation.

---

[3] A desirable property of a monetary instrument such as BitCoin is that it holds its value over short-medium periods of time in order not to create distortion when used as a medium of exchange in transaction. Large price movements alter the purchasing power potentially causing costs and risk to firms and consumers using it as a medium of exchange in transaction of goods and services.

[4] BitCoins are created in a 'mining' process, in which computer network participants, i.e. users who provide their computing power, verify and record payments into a public ledger called blockchain. In return for this service they receive transaction fees and newly minted BitCoins.

## 3 Conceptual framework

According to previous studies (Buchholz et al. 2012; Kristoufek 2013; van Wijk 2013), BitCoin price is determined by three key factors (i) supply-demand interactions of BitCoin, (ii) BitCoin's attractiveness for investors, and (iii) global macroeconomic and financial developments.

### 3.1 Supply-demand interactions

According to Buchholz et al. (2012), one of the key drivers of BitCoin price is the interaction between BitCoin supply and demand on the BitCoin market. The demand for BitCoin is primarily driven by its value as a medium of exchange (i.e. by value in future exchange). The supply is given by the stock of BitCoins in circulation, which is publicly known and is predefined in the long run.

The impact of supply-demand interactions on BitCoin price formation can be derived from a modified version of Barro's (1979) model for gold standard.[5] For the sake of comparability, we denominate the stock of money base of BitCoins in a standard government controlled fiat currency such as dollars.[6] As in Barro, we assume that firms need to convert BitCoins into dollars or other currencies, as they operate in economies using dollars or other currencies for purchase production factors.[7]

Suppose that $B$ represents the total stock of BitCoins in circulation and $P_B$ denotes the exchange rate of BitCoin (i.e. dollar per unit of BitCoin), then the total BitCoin money supply, $M^S$, is given by:

(1) $$M^S = P^B B$$

The demand for circulating BitCoins in dollar denomination, $M^D$, is assumed to depend on the general price level of goods and services, $P$, the size of BitCoin economy, $Y$, and the velocity of BitCoin circulation, $V$. The BitCoin's velocity, $V$, measures the frequency at which one unit of BitCoin is used for purchase of goods and services, and it depends on the opportunity cost for holding it (inflation, opportunity interest rate).

(2) $$M^D = \frac{PY}{V}$$

The equilibrium between BitCoin supply (1) and BitCoin demand (2) implies the following equilibrium price relationship:

---

[5] Barro (1979) developed model for gold standard. The key difference between the gold standard and BitCoin is that the demand for BitCoin is driven by its value in future exchange, whereas the demand for commodity currency is driven by both its intrinsic value and its value in future exchange. A second main difference is in the supply behaviour. The supply of commodity currency is endogenous; it responds to changes in production technology (e.g. mining technology for gold) and returns. Under current system, BitCoin supply is exogenous as it is predefined by the software algorithm.
[6] Note that goods and services are traded using dollars or other precious metals and not BitCoins.
[7] If all global transactions would be executed in BitCoins, then the monetary base would be fully BitCoin denominated and, in principle, its conversion to other currency would not be necessary.

(3) $$P^B = \frac{PY}{VB}$$

In perfect markets the price equilibrium given by equation (3) implies that the price of BitCoin decreases with the velocity and the stock of BitCoins, but increases with the size of BitCoin economy and the price level. Applying a logarithmic transformation to equation (3) and denoting variables in natural logs in lowercase, we can rewrite equation (3) into an empirically estimable model of BitCoin price:

(4) $$p_t^B = \beta_0 + \beta_1 p_t + \beta_2 y_t + \beta_3 v_t + \beta_4 b_t + \epsilon_t$$

where $\epsilon_t$ is error term. According to the underlying theoretical framework of Barro (1979), we expect that $\beta_1$ and $\beta_2$ would be positive, whereas $\beta_3$ and $\beta_4$ would be negative.

**3.2 BitCoin's attractiveness for investors**

BitCoin has been created relatively recently, particularly, when compared to other investment goods, such as gold. As a result, there are several important factors, which affect the behaviour of BitCoin investors in addition to the traditional ones (Barber et al. 2012; Buchholz et al. 2012; Kristoufek 2013; van Wijk 2013).

First, BitCoin price may be affected by risk and uncertainty of the BitCoin system. Given that BitCoin is a fiat currency and thus intrinsically worthless, it does not have an underlying value derived from consumption or its use in production process (such as gold). The value of fiat currency is based on trust that it will be valuable and accepted as a medium of exchange also in the future (Greco 2001).[8] The expectations about trust and acceptance are particularly relevant for BitCoin, which being a relatively new currency is in the phase of establishing its market share by building credibility among potential users.

Second, being a digital currency, BitCoin is more vulnerable to cyber-attacks, which can easily destabilise the whole BitCoin system and thus cause more volatile price responses. Such attacks have been occurring over the whole lifespan of BitCoin (Barber et al. 2012; Moore and Christin 2013). Moore and Christin (2013) examined 40 BitCoin exchanges and found that 18 have closed down due to cyber-attacks. For example, MtGox, once the world's biggest BitCoin exchange, collapsed in February 2014 due to a cyber-attack which allegedly led to a loss of 850 thousand BitCoins.

Third, investors behaviour and hence BitCoin's price is also determined by transactions costs for potential investors. According to Gervais, Kaniel, and Mingelgrin (2001), Grullon, Kanatas, and Weston (2004) and Barber and Odean (2008) the preferences of new investors' decision may be distorted by the effect of attention (e.g. attention in the news media) in the presence of many alternative investment choices and search costs. The attention-driven investment behaviour results from the costs associated with searching for information for potential investment opportunities available on the market, such as on the stock exchange. Investment opportunities under attention of news media may be preferred by new investors, because they reduce search costs thus triggering high price responses. Indeed, Lee (2014) finds such behaviour for BitCoin, whereby the alteration of positive and negative news generated high price cycles. This implies that the attention-driven investment behaviour can affect BitCoin price either positively or negatively, depending on the type of news that dominate in the media at a given point of time.

---

[8] Given that people consider a currency valuable if they expect others to do so, for a decentralised currency, such as BitCoin, that trust depends on a belief that the rules of the currency will be stable over time.

In order to account for BitCoin's attractiveness for investors in the BitCoin price formation, we extend equation (4) as follows:

(5) $$p_t^B = \beta_0 + \beta_1 p_t + \beta_2 y_t + \beta_3 v_t + \beta_4 b_t + \beta_5 a_t + \varepsilon_t$$

where $a_t$ is captures BitCoin's attractiveness for investors. According to the previous studies, coefficient $\beta_5$ can be either negative or positive.

**3.3 Macroeconomic and financial developments**

Van Wijk (2013) stresses the role of global macroeconomic and financial development, captured by variables such as stock exchange indices, exchange rates, and oil prices measures in determining BitCoin price. The impact of macroeconomic and financial indicators on BitCoin price may work through several channels. For example, stock exchange indices may reflect the general macroeconomic and financial developments of the global economy. Favourable macroeconomic and financial developments may stimulate the use of BitCoin in trade and exchanges and thus strengthen its demand which may have positive impact on BitCoin price.

Inflation and price indices are other important indicators of macroeconomic and financial developments. According to Krugman and Obstfeld (2003) and Palombizio and Morris (2012), oil price is one of the main sources of demand and cost pressures and provides an early indication of inflation development. Thus, if the price of oil signals potential changes in the general price level, this may lead to a depreciation of BitCoin. Also the exchange rate may reflect inflation development and thus impact positively BitCoin price as indicated by equation (3).

According to Dimitrova (2005), there could be also negative relation between BitCoin price and macro financial indicators. A decline in stock prices induces foreign investors to sell the financial assets they hold. This leads to a depreciation of the respective currency, but may stimulate BitCoin price if investors substitute investment in stock for investment in BitCoin. Generally, investors' return on stock exchange may capture opportunity costs of investing in BitCoin. Hence, stock exchange indices are expected to be positively related with BitCoin price.

In order to account for macroeconomic and financial developments in the BitCoin price formation, we extend equation (5) as follows:

(6) $$p_t^B = \beta_0 + \beta_1 p_t + \beta_2 y_t + \beta_3 v_t + \beta_4 b_t + \beta_5 a_t + \beta_6 m_t + \varepsilon_t$$

where $m_t$ is captures macroeconomic and financial indicators. According to the previous studies, we expect $\beta_6$ to be either positive or negative.

## 4 Data

We use the following proxies to capture the supply-demand fundamentals suggested by the price relationship (3). We use data for BitCoin price, $P^B$, denominated in US dollars (*mkpru*). We use a historical number of total BitCoins (*totbc*) which have been mined to account for the total stock of BitCoins in circulation $B$. We use two alternative proxies for BitCoin economy, $Y$: the total number of unique BitCoin transactions per day (*ntran*), and the number of unique BitCoin addresses used per day (*naddu*). Following Matonis (2012), we proxy the monetary BitCoin's velocity, $V$, by BitCoin days destroyed for any given transaction, *bcdde*. This variable is calculated by taking the number of BitCoin in a transaction and multiplying it by the number of days since those coins were last spent. All these data are extracted from *quandl.com*. To measure the price level of global economy, $P$, we use exchange rate between the U.S. dollar and the Euro (*exrate*) extracted from the European Central Bank.

In order to capture BitCoin's attractiveness for investors, $a$, following Kristoufek (2013), we use the volume of daily BitCoin views on Wikipedia, *wiki_views*, which measures investors' faith in BitCoin.[9] According to Kristoufek (2013), the frequency of searches related to the digital currency is a good measure of potential investors' interest in the currency. Piskorec et al. (2014) argue that this measure may also capture speculative behaviour of investors. The online search queries, such as Wikipedia views, may measure investors' interest in BitCoin, as it captures the information's demand about the currency. We use also variable of the number of new members (*new_members*) and new posts (*new_posts*) extracted from *bitcointalk.org*. As explained above, the variable *new_members* captures the size of the BitCoin economy but also attention-driven investment behaviour of new BitCoin members. The variable *new_posts* captures the effect of trust and/or uncertainty, as it represents the intensity of discussions among members.

To account for global macroeconomic and financial indicators, $m$, we follow van Wijk (2013) and we use oil price, *oil_price*, and the Dow Jones stock market index, *DJ*.[10] The oil prices are from the US Energy Information Administration and Dow Jones index is extracted from the Federal Research Bank of St. Louis.

---

[9] Kristoufek (2013) used also queries of BitCoin on Google Trends to measure investor faith/sentiment in BitCoin. These data are available only on weekly bases. Since we use daily data we do not use this proxy in our estimations.

[10] The Dow Jones Index is an industrial average that captures 30 major corporations on either the NYSE or the NASDAQ.

## 5 Econometric approach

The conceptual analyses in section 3 suggest that BitCoin price and explanatory variables considered in the analysis are mutually interdependent. The estimation of non-linear interdependencies among interdependent time series in presence of mutually cointegrated variables is subject to the endogeneity problem (Lütkepohl and Krätzig 2004). To circumvent the problem of endogeneity, we follow the general approach in the literature to analyse the causality between endogenous time-series and specify a Vector Auto-Regressive (VAR) model (Lütkepohl and Krätzig 2004).

According to Engle and Granger (1987), regressions of interdependent and non-stationary time series may lead to spurious results. In order to avoid spurious regression, it is important to test the properties of the time series involved. Therefore, in the first step, the stationarity of time series is determined, for which we use two unit root tests: the augmented Dickey-Fuller (ADF) test and the Phillips-Perron (PP) test. The number of lags that we use for each dependent variable is determined by the Akaike Information Criterion (AIC). If two individual time series are not stationary, their combination may be stationary (Engle and Granger 1987). In this special case, the time series are considered to be cointegrated, implying that there exists a long-run equilibrium relationship between them.

In the second step, we employ the Johansen's cointegration method to examine the long-term relationship between the price series. The number of cointegrating vectors is determined by the maximum eigenvalue test and the trace test. Both tests use eigenvalues to compute the associated test statistics. We follow the Pantula principle to determine whether a time trend and a constant term should be included in the model.

In the third step, we estimate a vector error correction model for those series that are cointegrated. It includes an error correction term indicating the speed of adjustment of any disequilibrium towards a long-term equilibrium state. Following Johansen and Juselius's (1990), we start with a vector autoregressive model and reformulate it into a vector error correction model:

(7)
$$Z_t = A_1 Z_{t-1} + ... + A_k Z_{t-k} + \varepsilon_t$$

(8)
$$\Delta Z_t = \sum_{i=1}^{k-1} \Gamma_i \Delta Z_{t-i} + \Pi Z_{t-1} + \varepsilon_t$$

where $Z_t$ is a vector of non-stationary variables, $A$ are matrices of different parameters, $t$ is time subscript, $k$ is the number of lags and $\varepsilon_t$ is the error term assumed to follow *i.i.d.* process with a zero mean and normally distributed $N(0, \sigma2)$ error structure. Equation (2) contains information on both short-run and long-run adjustments to changes in $Z_t$ via the estimates of $\Gamma_i$ and $\Pi$, respectively. $\Pi$ can be decomposed as $\Pi=\alpha\beta'$, where $\alpha$ represents the speed of adjustment to disequilibrium and $\beta$ represents the long-run relationships between variables (Johansen and Juselius's 1990).

As usual, in order to ensure the adequacy of the estimated models, we implement a series of specification tests: Lagrange-multiplier (LM) test for autocorrelation in the residuals; Jarque-Berra test to check if the residuals in the VEC are normally distributed and the test of stability of the model.

## 6 Results

Following the theoretical hypothesis, we estimate four sets of econometric models of BitCoin price (differences in specifications between the estimated models are reported in Table 1). Models 1.1 to 1.5 capture BitCoin's supply-demand interactions and their impact on BitCoin price. Model 2.1 estimates the impact of BitCoin's attractiveness for investors of buying/selling BitCoins. Model 3.1 estimates the impact of global macroeconomic and financial developments. Models 4.1 to 4.9 interact the above three components.

Given that several variables are highly correlated, we estimate alternative model specifications (Table 1) by sequentially replacing those variables that are highly correlated.[11] We found a particularly high correlation (corr. >0.8) for the total stock of BitCoin and the size of the BitCoin economy (*totbc – ntran*, *totbc – naddu*, *naddu – ntran*), the Dow Jones Index and the total stock of BitCoins, and the size of BitCoin economy (*totbc – dj*, *naddu – dj*) and the Wikipedia views with the Dow Jones Index and the size of BitCoin economy (*wiki_views – dj*, *wiki_views – naddu*) (Table 2).

The estimation results are reported in Table 3, Table 4, Table 5 and Table 6. Whereas Table 3 and Table 4 report the short-run impacts, Table 5 and Table 6 show the long-run impact of different determinants on BitCoin price. According to the results reported in Table 3 and Table 4, a number of variables have statistically significant short-run effect on BitCoin price adjustments. In particular, this is the case for own price effects, the stock of total BitCoins, *totbc*, BitCoin days destroyed, *bcdde*, and Wikipedia views, *wiki_views*. The short-run effects represent the short-run dynamics of variables in the cointegrated system. It describes how the time series react when the long-run equilibrium is distorted.

According to the results reported in Table 5 and Table 6, the long-run relationship between BitCoin price and different variables considered in the estimated models is stronger than the short-run impact. The first major observation arising from our estimates is that the supply-demand fundamentals have a strong impact on BitCoin price. The demand side variables (e.g. *bcdde*, *naddu*) appear to exert a stronger impact on BitCoin price than the supply side drivers (e.g. *totbc*). According to the results reported in Table 5, an increase in the stock of BitCoins (*totbc*) leads to a decrease in BitCoin price (model 1.2), whereas an increase in the size of the BitCoin economy (*naddu*) and its velocity (*bcdde*) lead to a higher price (models 1.1, 1.2, 1.3, 1.4, 1.5, 4.3, 4.4, 4.7). Contrary to our expectations, the alternative variable that captures the size of the BitCoin economy (*ntran*) has negative impact on BitCoin price in models 1.1 and 1.5. However, this variable is not significant in the more general models (models 4.1-4.9).

---

[11] Note that we have tested for the stationarity of the data series using augmented Dickey Fuller (ADF) and Phillips Perron (PP) tests. The lags of the dependent variable in the tests were determined by Akaike Information Criterion (AIC). Both tests show that all the time series are non-stationary in levels but stationary in first differences (results of the tests are available upon request from authors).

Although, the sign of the estimated coefficients for supply-demand fundamentals of BitCoin is in line with the theoretical predictions (except for *ntran* in models 1.1 and 1.5), the statistical significance and magnitude of the estimated coefficients decreases in most models, when accounting for the impact of BitCoin's attractiveness for investors and global macroeconomic and financial developments (models 4.1 to 4.9 in Table 6). The supply-demand variables are statistically significant in models 4.3, 4.4 and 4.7, but have a considerably lower magnitude of the estimated impact than in models 1.1 to 1.5, which capture only BitCoin's supply-demand fundamentals. This could be explained by the fact that part of the BitCoin's price variation explained by the supply-demand variables is absorbed by other variables in more general specifications (models 4.1 to 4.9).

The strongest and statistically most significant impact on BitCoin price is estimated for variables capturing the impact of BitCoin's attractiveness for investors: *wiki_views*, *new_members* and *new_posts* (models 2.1 and models 4.1 to 4.9). The variable *new_members* has negative impact on BitCoin price, implying that attention-driven investment behaviour of new members dominates. The variable *new_posts* has positive impact on BitCoin price, reflecting an increasing acceptance and trust of BitCoin captured by the intensity of discussion between BitCoin users. This may reflect declining transaction costs and uncertainty for investors, which increases investment demand of BitCoins and hence it's price.

Consistent with Kristoufek (2013), Wikipedia views have a statistically significant impact on BitCoin price. This variable is significant and has positive impact in all models in which it (except for model 4.8). However, the interpretation of this variable is not straightforward, as it may capture various effects. On the one hand, this may reflect speculative behaviour of investors. Kristoufek (2013) argues that, since the BitCoin fundamentals allowing for setting a ''fair'' price are missing, its price is driven by the investors' faith in the future growth and is dominated by short-term investors, trend chasers, noise traders and speculators. On the other hand, Wikipedia views may measure investors' interest in BitCoin, as it captures information's demand about the currency (Piskorec et al. 2014). It may reflect changes in the knowledge about BitCoin between users, thus leading to a higher acceptance and demand for it. Important is that the type of people searching information about BitCoin on Wikipedia are likely to be new BitCoin users/investors, because Wikipedia contains rather general information about BitCoin, which is known by incumbent investors or advanced BitCoin users. If these last two arguments hold, then the estimated Wikipedia effect represents the impact of the demand side of the BitCoin economy as given by variable *Y* in equation (3) not necessarily capturing only speculative behaviour of investors.

Our findings suggest that, in contrast to previous studies (i.e. van Wijk 2013), macro-financial indicators such as the Dow Jones Index, exchange rate and oil price do not significantly affect BitCoin price in the long-run. Only in Model 3.1 all macro and financial variables (*dj*, *oil_price* and *exrate*) are statistically significant (Table 5). This is in line with the estimates of van Wijk (2013), who also finds statistically significant impact of macro-financial variables on BitCoin price. However, van Wijk (2013) does not account for BitCoin's market fundamentals or BitCoin's attractiveness for investors. When these factors are taken in consideration (models 4.1 to 4.9), their impact decreases considerably in all models (except for model 4.1) (Table 6).

# 7 Conclusions

Due to a growing market share of BitCoin, a rapidly increasing price of BitCoin and its high price volatility, there is an increasing interest among users and academics in understanding the BitCoin system in general and its price formation in particular. This paper attempts to shed light on drivers that determine BitCoin price in the short- and long-run. The paper analyses the relationship between BitCoin price and supply-demand fundamentals of BitCoins alongside the global macro-financial indicators and BitCoin's attractiveness for investors. We employ a VAR estimation approach and use daily data for the period 2009-2014 to identify the causal effects between BitCoin's price and its determinants.

Our empirical analyses confirm that BitCoin market fundamentals have an important impact on BitCoin price, implying that, to a large extent, the formation of BitCoin price can be explained in a standard economic model of currency price formation. Supply and demand drivers have an important impact on the BitCoin price formation and thus are among the key factors in determining its stability. In particular, the demand-side drivers, such as the size of the BitCoin economy and the velocity of BitCoin circulation, have the strongest impact on BitCoin price. Hence, given that BitCoin supply is exogenous, the development of the demand side drivers will be among the key determinants of BitCoin price also in the future.

Second, we cannot reject the hypothesis that speculations are also affecting BitCoin price. The statistically significant impact of Wikipedia views on BitCoin price could be an evidence of speculative short-run behaviour of investors, or it may capture the expansion of the demand side of the BitCoin economy. Additionally, we find that also new posts impact BitCoin price positively, which may be a result of an increased trust among users. As such, speculative trading of BitCoins is not necessarily an undesirable activity, as it may generate benefits in terms of absorbing excess risk from risk adverse participants and providing liquidity on the market. A negative side of short-run investors' speculative behaviour is that it may increase price volatility and price bubbles. The success of BitCoin thus also hitches on its ability to reduce the potential negative implications of speculations and expand the use of BitCoin in trade and commerce.

Finally, our estimates do not support previous findings that macro-financial indicators may be driving BitCoin price. In fact, once we control for supply-demand variables and BitCoin's attractiveness for investors linked to BitCoin, the relevance of macro-financial indicators becomes statistically insignificant.

Kroll, J., I. Davey, and E. Felten (2013)."The Economics of BitCoin Mining, or BitCoin in the Presence of Adversaries." WEIS 2013, http://weis2013.econinfosec.org/papers/KrollDaveyFeltenWEIS2013.pdf

Lee, T.B. (2014). "These four charts suggest that BitCoin will stabilize in the future." *Washington Post*, http://www.washingtonpost.com/blogs/the-switch/wp/2014/02/03/these-four-charts-suggest-that-bitcoin-will-stabilize-in-the-future/

Lütkepohl, H., Krätzig, M. (2004). Applied Time Series Econometrics, Cambridge University Press.

Matonis, J. (2012). "Top 10 BitCoin Statistics." *Forbes* 7/31/2012, http://www.forbes.com/sites/jonmatonis/2012/07/31/top-10-bitcoin-statistics

Moore, T. and N. Christin (2013). "Beware the Middleman: Empirical Analysis of BitCoin-Exchange Risk." *Financial Cryptography and Data Security* 7859: 25-33.

Murphy, R.P. (2013). "The Economics of BitCoin." Library Economic Liberty, http://www.econlib.org/library/Columns/y2013/MurphyBitCoin.html

Palombizio E. and I. Morris. (2012). "Forecasting Exchange Rates using Leading Economic Indicators." *Open Access Scientific Reports* 1(8): 1-6.

Piskorec, P., N. Antulov-Fantulin, P. K. Novak, I. Mozetic, M. Grcar, I. Vodenska, and T. Šmuc (2014). "News Cohesiveness: an Indicator of Systemic Risk in Financial Markets." arXiv:1402.3483v1 [cs.SI], http://arxiv.org/pdf/1402.3483v1.pdf

van Wijk, D. (2013). "What can be expected from the BitCoin?" Working Paper No. 345986, Erasmus Rotterdam Universiteit.

**Table 1: Specification of the empirically estimated models**

| | | M 1.1 | M 1.2 | M 1.3 | M 1.4 | M 1.5 | M 2.1 | M 3.1 | M 4.1 | M 4.2 | M 4.3 | M 4.4 | M 4.5 | M 4.6 | M 4.7 | M 4.8 | M 4.9 |
|---|---|---|---|---|---|---|---|---|---|---|---|---|---|---|---|---|---|
| Supply-demand variables | totbc | x | x | x | | | | | | x | x | | x | | | | |
| | ntran | x | | | | x | | | | | | | | | | x | x |
| | naddu | | x | | x | | | | | x | x | x | | | x | | |
| | bcdde | x | x | x | x | x | | | | x | x | x | x | x | x | x | x |
| | exrate | x | x | x | x | x | | x | x | x | | | | | | | |
| BitCoin's attractiveness for investors | wiki_views | | | | | | x | | x | x | x | x | | x | | x | |
| | new_member | | | | | | x | | | x | x | | x | x | | x | x |
| | new_posts | | | | | | x | | | x | x | x | x | x | x | x | x |
| Macro-financial developments | dj | | | | | | | x | x | x | x | | | | | | x |
| | oil_price | | | | | | | x | x | x | | | | | | | |

**Table 2: Correlation coefficients**

|             | ntran | totbc | bcdde | dj   | wiki_views | naddu | oil_price | ex_rate | new_posts | new_members |
|-------------|-------|-------|-------|------|------------|-------|-----------|---------|-----------|-------------|
| ntran       | 1     |       |       |      |            |       |           |         |           |             |
| totbc       | **0.92** | 1  |       |      |            |       |           |         |           |             |
| bcdde       | 0.45  | 0.41  | 1     |      |            |       |           |         |           |             |
| dj          | 0.76  | **0.92** | 0.38 | 1 |            |       |           |         |           |             |
| wiki_views  | 0.71  | 0.79  | 0.51  | **0.83** | 1     |       |           |         |           |             |
| naddu       | **0.92** | **0.95** | 0.47 | **0.91** | **0.86** | 1 |           |         |           |             |
| oil_price   | -0.21 | 0.03  | -0.06 | 0.29 | 0.07       | -0.01 | 1         |         |           |             |
| ex_rate     | 0.19  | 0.44  | 0.19  | 0.59 | 0.50       | 0.42  | 0.50      | 1       |           |             |
| new_posts   | 0.59  | 0.68  | 0.34  | 0.75 | 0.68       | 0.74  | 0.11      | 0.36    | 1         |             |
| new_members | 0.40  | 0.45  | 0.24  | 0.51 | 0.59       | 0.53  | 0.01      | 0.25    | 0.66      | 1           |

**Table 3: Short-run effects on BitCoin Price for model sets 1, 2 and 3**

| | M 1.1 | M 1.2 | M 1.3 | M 1.4 | M 1.5 | M 2.1 | M 3.1 |
|---|---|---|---|---|---|---|---|
| LD.mkpru | 0.147*** | 0.136*** | 0.149*** | 0.135*** | 0.145*** | 0.147*** | 0.143*** |
| L2D.mkpru | -0.017 | -0.021 | -0.017 | -0.020 | -0.014 | -0.020 | - |
| L3D.mkpru | -0.033 | -0.027 | -0.037 | -0.028 | -0.025 | -0.032 | - |
| L4D.mkpru | 0.054* | - | 0.051 | - | - | - | - |
| LD.totbc | -17.31 | -11.940 | -21.540 | - | - | - | - |
| L2D.totbc | 38.51 | 44.880* | 36.500 | - | - | - | - |
| L3D.totbc | -43.68* | -20.140 | -48.360** | - | - | - | - |
| L4D.totbc | 41.60** | - | 39.190* | - | - | - | - |
| LD.ntran | 0.001 | - | - | - | 0.000 | - | - |
| L2D.ntran | -0.002 | - | - | - | 0.003 | - | - |
| L3D.ntran | -0.016 | - | - | - | -0.018 | - | - |
| L4D.ntran | -0.011 | - | - | - | - | - | - |
| LD.naddu | - | 0.015 | - | 0.011 | - | - | - |
| L2D.naddu | - | 0.005 | - | 0.007 | - | - | - |
| L3D.naddu | - | -0.022 | - | -0.023 | - | - | - |
| L4D.naddu | - | - | - | - | - | - | - |
| LD.bcdde | -0.009** | -0.009** | -0.011*** | -0.009*** | -0.008** | - | - |
| L2D.bcdde | -0.007* | -0.006* | -0.008** | -0.006** | -0.006* | - | - |
| L3D.bcdde | -0.005 | -0.003 | -0.007* | -0.003 | -0.003 | - | - |
| L4D.bcdde | -0.002 | - | -0.003 | - | - | - | - |
| LD.exrate | -0.382 | -0.391 | -0.417 | -0.404 | -0.426 | - | -0.630 |
| L2D.exrate | 0.369 | 0.429 | 0.327 | 0.429 | 0.359 | - | - |
| L3D.exrate | 0.175 | 0.292 | 0.152 | 0.280 | 0.222 | - | - |
| L4D.exrate | -0.278 | - | -0.306 | - | - | - | - |
| LD.wiki_views | - | - | - | - | - | -0.005 | - |
| L2D.wiki_views | - | - | - | - | - | -0.012* | - |
| L3D.wiki_views | - | - | - | - | - | -0.015** | - |
| L4D.wiki_views | - | - | - | - | - | - | - |
| LD.new_members | - | - | - | - | - | 0.004 | - |
| L2D.new_members | - | - | - | - | - | 0.005 | - |
| L3D.new_members | - | - | - | - | - | -0.001 | - |
| L4D.new_members | - | - | - | - | - | - | - |
| LD.new_posts | - | - | - | - | - | -0.007 | - |
| L2D.new_posts | - | - | - | - | - | -0.002 | - |
| L3D.new_posts | - | - | - | - | - | 0.005 | - |
| L4D.new_posts | - | - | - | - | - | - | - |
| LD.dj | - | - | - | - | - | - | -0.046 |
| L2D.dj | - | - | - | - | - | - | - |
| L3D.dj | - | - | - | - | - | - | - |
| L4D.dj | - | - | - | - | - | - | - |
| LD.oilprice | - | - | - | - | - | - | 0.178 |
| L2D.oilprice | - | - | - | - | - | - | - |
| L3D.oilprice | - | - | - | - | - | - | - |
| L4D.oilprice | - | - | - | - | - | - | - |
| constant | - | - | 0.000 | - | - | - | - |

Notes: *** significant at 1% level, ** significant at 5% level, * significant at 10% level. "-" indicates either absence of a variable in the respective model or the coefficient is not significantly different from zero.

**Table 4: Short-run effects on BitCoin Price for general models**

| | M 4.1 | M 4.2 | M 4.3 | M 4.4 | M 4.5 | M 4.6 | M 4.7 | M 4.8 | M 4.9 |
|---|---|---|---|---|---|---|---|---|---|
| LD.mkpru | 0.144*** | 0.136*** | 0.147*** | 0.146*** | 0.145*** | 0.149*** | 0.143*** | 0.148*** | 0.147*** |
| L2D.mkpru | - | -0.032 | -0.017 | -0.026 | -0.027 | -0.024 | -0.020 | -0.016 | -0.017 |
| L3D.mkpru | - | - | -0.021 | - | - | - | -0.022 | -0.033 | -0.028 |
| L4D.mkpru | - | - | - | - | - | - | - | 0.056* | 0.054* |
| LD.totbc | - | -20.930 | -15.524 | - | -16.637 | - | - | - | - |
| L2D.totbc | - | 26.010 | 44.028*** | - | 31.547 | - | - | - | - |
| L3D.totbc | - | - | -32.434 | - | - | - | - | - | - |
| L4D.totbc | - | - | - | - | - | - | - | - | - |
| LD.ntran | - | - | - | - | - | - | - | 0.001 | 0.001 |
| L2D.ntran | - | - | - | - | - | - | - | 0.006 | 0.002 |
| L3D.ntran | - | - | - | - | - | - | - | -0.018 | -0.020 |
| L4D.ntran | - | - | - | - | - | - | - | -0.009 | -0.007 |
| LD.naddu | - | 0.024 | 0.001 | 0.009 | - | - | 0.006 | - | - |
| L2D.naddu | - | 0.014 | -0.002 | 0.007 | - | - | 0.001 | - | - |
| L3D.naddu | - | - | -0.022 | - | - | - | -0.026* | - | - |
| L4D.naddu | - | - | - | - | - | - | - | - | - |
| LD.bcdde | - | -0.003 | -0.004 | -0.005 | -0.005 | -0.004 | -0.009** | -0.006 | -0.009** |
| L2D.bcdde | - | -0.002 | -0.002 | -0.004 | -0.003 | -0.003 | -0.006* | -0.005 | -0.008* |
| L3D.bcdde | - | - | -0.001 | - | - | - | -0.003 | -0.003 | -0.005 |
| L4D.bcdde | - | - | - | - | - | - | - | -0.002 | -0.003 |
| LD.exrate | -0.644 | -0.488 | - | - | - | - | - | - | - |
| L2D.exrate | - | 0.370 | - | - | - | - | - | - | - |
| L3D.exrate | - | - | - | - | - | - | - | - | - |
| L4D.exrate | - | - | - | - | - | - | - | - | - |
| LD.wiki_views | -0.002 | -0.003 | -0.007 | 0.001 | - | 0.001 | - | -0.004 | - |
| L2D.wiki_views | - | -0.010 | -0.013* | -0.009 | - | -0.008 | - | -0.011 | - |
| L3D.wiki_views | - | - | -0.014** | - | - | - | - | -0.012* | - |
| L4D.wiki_views | - | - | - | - | - | - | - | 0.004 | - |
| LD.new_members | - | 0.003 | 0.003 | 0.003 | 0.003 | 0.004 | - | 0.004 | 0.002 |
| L2D.new_members | - | 0.005 | 0.005 | 0.004 | 0.004 | 0.005 | - | 0.005 | 0.002 |
| L3D.new_members | - | - | -0.001 | - | - | - | - | -0.001 | -0.003 |
| L4D.new_members | - | - | - | - | - | - | - | 0.001 | 0.000 |
| LD.new_posts | - | -0.006 | -0.005 | -0.006 | -0.005 | -0.007 | - | -0.007 | -0.006 |
| L2D.new_posts | - | -0.003 | -0.001 | -0.003 | -0.003 | -0.003 | -0.003 | -0.002 | -0.001 |
| L3D.new_posts | - | - | 0.006 | - | - | - | 0.001 | 0.004 | 0.004 |
| L4D.new_posts | - | - | - | - | - | - | 0.004 | -0.001 | -0.001 |
| LD.dj | -0.057 | -0.143 | 0.009 | - | - | - | - | - | 0.085 |
| L2D.dj | - | 0.088 | 0.008 | - | - | - | - | - | 0.010 |
| L3D.dj | - | - | -0.226 | - | - | - | - | - | -0.209 |
| L4D.dj | - | - | - | - | - | - | - | - | 0.094 |
| LD.oilprice | 0.181 | 0.212 | - | - | - | - | - | - | - |
| L2D.oilprice | - | -0.176 | - | - | - | - | - | - | - |
| L3D.oilprice | - | - | - | - | - | - | - | - | - |
| L4D.oilprice | - | - | - | - | - | - | - | - | - |
| constant | - | - | - | - | - | - | - | - | - |

Notes: *** significant at 1% level, ** significant at 5% level, * significant at 10% level. "-" indicates either absence of a variable in the respective model or the coefficient is not significantly different from zero.

**Table 5: Long-run effects on BitCoin Price for model sets 1, 2 and 3**

|              | M 1.1     | M 1.2     | M 1.3     | M 1.4    | M 1.5      | M 2.1     | M 3.1      |
|-------------:|-----------|-----------|-----------|----------|------------|-----------|------------|
| totbc        | -4.2      | -5.96***  | -         | -        | -          | -         | -          |
| ntran        | -3.99**   | -         | -         | -        | -3.42***   | -         | -          |
| naddu        | -         | 3.17***   | -         | -        | -          | -         | -          |
| bcdde        | 11.71***  | -         | 5.07***   | 5.40***  | 10.84***   | -         | -          |
| exrate       | -         | -         | 12.69     | -        | 0.43       | -         | 16.04***   |
| wiki_views   | -         | -         | -         | -        | -          | 1.94***   | -          |
| new_members  | -         | -         | -         | -        | -          | -         | -          |
| new_posts    | -         | -         | -         | -        | -          | -         | -          |
| dj           | -         | -         | -         | -        | -          | -         | 16.10***   |
| oil_price    | -         | -         | -         | -        | -          | -         | - 4.52***  |
| constant     | -62.14    | -66.73**  | -77.35*** | -77.79***| -124.98*** | -13.66*** | -133.79*** |

Notes: *** significant at 1% level, ** significant at 5% level, * significant at 10% level
"-" indicates either absence of a variable in the respective model or the coefficient is not significantly different from zero.

**Table 6: Long-run effects on BitCoin Price for general models**

|             | M 4.1     | M 4.2     | M 4.3     | M 4.4     | M 4.5     | M 4.6     | M 4.7     | M 4.8     | M 4.9     |
|------------:|-----------|-----------|-----------|-----------|-----------|-----------|-----------|-----------|-----------|
| totbc       | -         | -         | -         | -         | -         | -         | -         | -         | -         |
| ntran       | -         | -         | -         | -         | -         | -         | -         | -         | -         |
| naddu       | -         | -         | 0.36      | 0.50***   | -         | -         | -         | -         | -         |
| bcdde       | -         | -         | 0.41***   | -         | -         | -         | 0.28**    | -         | -         |
| exrate      | 0.63      | -         | -         | -         | -         | -         | -         | -         | -         |
| wiki_views  | 1.38***   | 0.89***   | 0.90***   | 1.78***   | -         | 1.93***   | -         | -         | -         |
| new_members | -         | -1.23***  | -0.26***  | -0.35***  | -         | -         | -         | -0.46***  | -0.35***  |
| new_posts   | -         | 2.33***   | 1.21***   | -         | 1.92***   | -         | 1.99***   | 2.55***   | 2.44***   |
| dj          | 4.79***   | -         | -         | -         | -         | -         | -         | -         | -         |
| oil_price   | -         | -         | -         | -         | -         | -         | -         | -         | -         |
| constant    | -54.45*** | -17.44**  | -15.64*** | -15.73*** | -11.89*** | -13.62*** | -17.23*** | -15.29*** | -14.96*** |

Notes: *** significant at 1% level, ** significant at 5% level, * significant at 10% level
"-" indicates either absence of a variable in the respective model or the coefficient is not significantly different from zero.

**Figure 1. BitCoin price development 2009-2014**

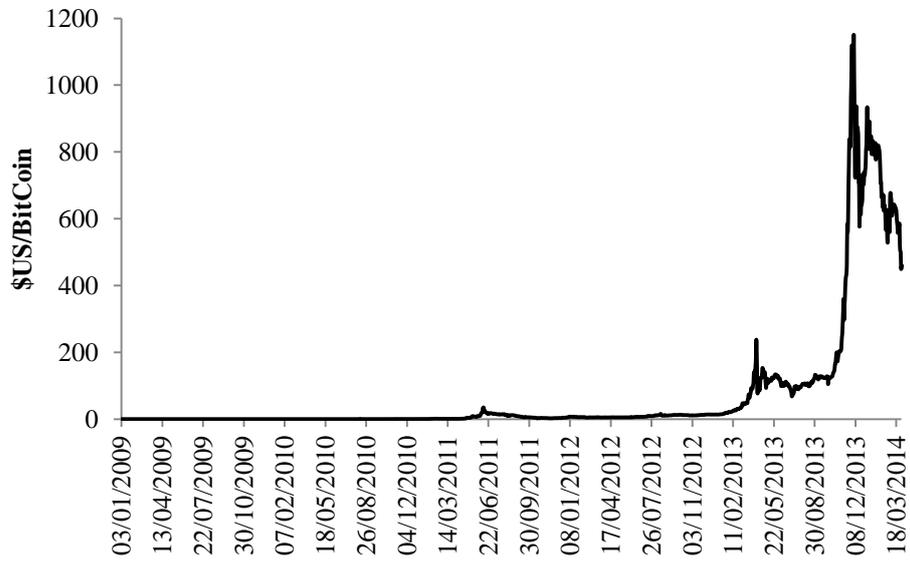

Source: http://www.quandl.com/markets/BitCoin

22